\documentclass[letterpaper, 10 pt, conference]{ieeeconf}  

\IEEEoverridecommandlockouts                              

\usepackage[dvipsnames]{xcolor}
\usepackage{graphics} 
\usepackage{epsfig} 
\usepackage{caption}
\usepackage{subcaption}
\usepackage{graphicx}
\usepackage{booktabs}
\let\labelindent\relax
\usepackage{nicematrix,enumitem}
\usepackage{hyperref}  
\usepackage{amssymb}
\usepackage{algorithm,algorithmic}

\hypersetup{
    colorlinks=true,
    linkcolor=blue,
    filecolor=magenta,      
    urlcolor=cyan,
    pdftitle={Overleaf Example},
    pdfpagemode=FullScreen,
    }

\title{\LARGE \bf
Learning When to See for Long-term Traffic Data Collection on Power-constrained Devices}

\author{Ruixuan Zhang$^1$, Wenyu Han$^1$, Zilin Bian$^1$, Kaan Ozbay$^{1}$\textsuperscript{,*}, Chen Feng$^{1}$\textsuperscript{,*}
\thanks{$^1$ Tandon School of Engineering, New York University, NY, USA, 10012. }%
\thanks{* Corresponding Authors emails: {\tt\small \{ko772, cfeng\}@nyu.edu}}%
}

\begin{document}

\maketitle
\thispagestyle{empty}
\pagestyle{empty}

\begin{abstract}

Collecting traffic data is crucial for transportation systems and urban planning, and is often more desirable through easy-to-deploy but power-constrained devices, due to the unavailability or high cost of power and network infrastructure. The limited power means an inevitable trade-off between data collection duration and accuracy/resolution.
We introduce a novel learning-based framework that strategically decides observation timings for battery-powered devices and reconstructs the full data stream from sparsely sampled observations, resulting in minimal performance loss and a significantly prolonged system lifetime. Our framework comprises a predictor, a controller, and an estimator. The predictor utilizes historical data to forecast future trends within a fixed time horizon. The controller uses the forecasts to determine the next optimal timing for data collection. Finally, the estimator reconstructs the complete data profile from the sampled observations. We evaluate the performance of the proposed method on PeMS data by an RNN (Recurrent Neural Network) predictor and estimator, and a DRQN (Deep Recurrent Q-Network) controller, and compare it against the baseline that uses Kalman filter and uniform sampling. The results indicate that our method outperforms the baseline, primarily due to the inclusion of more representative data points in the profile, resulting in an overall 10\% improvement in estimation accuracy. Source code will be publicly available.

\end{abstract}

\section{Introduction}

Over the past few decades, Intelligent Transportation Systems (ITS) have experienced significant growth and advancement. The remarkable achievements of ITS can be largely attributed to the development of infrastructure and hardware for accessing, collecting, and processing real-world data. Within this context, surveillance cameras have emerged as a vital and extensively employed element of modern ITS, particularly due to the rapid progress in video processing and communication technology. Leveraging the valuable insights ingrained within high-dimensional image data, the utilization of collected video-based data presents immense possibilities for traffic analysis and encompasses various applications, including the detection of hazardous traffic scenarios such as collisions \cite{ijjina2019computer} and deteriorating road conditions \cite{takeuchi2012distinction}, as well as traffic counting \cite{dinh2021towards} and traffic state estimation \cite{li2013traffic}. Of greater significance, video-based data goes beyond extracting vehicle-related information and allows for capturing data regarding road users like pedestrians and cyclists, which traditional vehicle-focused sensors are difficult to provide~\cite{kothuri2017bicycle}. This opens up possibilities for various applications residing in crowd management and public safety, such as social distancing analysis during COVID-19 \cite{zuo2021reference}.

Among the many effective traffic data-collection systems that employ surveillance cameras, the Caltrans PeMS\footnote{\url{http://pems.dot.ca.gov}} and the New York City traffic sensing program\footnote{\url{https://webcams.nyctmc.org/map}} are two notable examples, where the Caltrans PeMS system gathers various real-time highway data, covering an extensive directional distance of over 41,000 miles, with more than 18,000 stations and 46,000 detectors, and the New York City traffic sensing program deploys over 2,000 surveillance cameras positioned at intersections across the city to monitor urban traffic conditions. Undoubtedly, these expansive data collection systems offer the research community and policymakers access to invaluable data that was previously unavailable. Nevertheless, we have identified three key challenges in the existing video-based data collection systems that limit the general public's accessibility and hinder the broader impact of leveraging these advanced technologies in ITS:

\begin{itemize}

    \item Expensive infrastructure limits the reach of existing surveillance systems, leaving certain underrepresented areas, like lower-income neighborhoods, being excluded from the monitoring and management efforts of transportation agencies. 
    
    \item The lack of spatial flexibility in infrastructure-based surveillance systems hampers response to urgent traffic monitoring demands (i.e., temporary work zones and special events) and hinders quick evaluation and adaptation for expansion requirements.
    
    \item The current data collection strategies that depend on infrastructure and necessitate a stable external power supply are impractical in situations where power accessibility is not guaranteed.
    
\end{itemize}


To address the limitations of hardware and the necessity for efficient data collection strategies, we present a learning-based data collection framework that empowers rapidly deployable lightweight devices to perform video-based data collection and extend their lifespan through the use of battery power. Specifically, the framework aims to address the inherent trade-off between extending the lifespan of the power-constrained devices and the resulting performance degradation caused by prolonged usage. The framework consists of three modularized components, namely prediction, control, and estimation. We endorse the concept of predictive control, wherein the DRQN controller leverages prediction results from the RNN predictor to actively determine the timing for the next data collection activity. Additionally, an RNN estimator is employed to refine the prediction results, aiming to minimize the disparity between the obtained and actual data profiles. The proposed framework is evaluated using a real-world traffic dataset, demonstrating its superiority over the baselines. To the best of our knowledge, there are currently no existing framework-level solutions for enabling long-term (self-maintained for at least a week) video data collection in power-constrained contexts.

\section{Related Work}
\label{section: II}

In this section, we review the existing work on enabling video processing and data collection under power-constrained schemes, which can be broadly classified into two categories: (1) reducing the computational load of neural networks for video processing, and (2) optimizing the sampling strategies.

\textbf{Efficient deep network backbone}. The progress in single-board computers has made it possible to run modern deep learning models on affordable devices equipped with dedicated GPU (graphics processing unit) or TPU (tensor processing unit), such as Nvidia Jetson Nano and Google Coral Dev Board. However, due to the limited onboard RAM and inferencing capabilities of these edge devices, most state-of-the-art image processing models (object detection, object tracking, crowd density estimation, etc.) cannot be directly deployed, and a significant accuracy drop is observed if limiting power supply \cite{alyamkin2019low}. Numerous efforts have been made to develop lightweight models for power-constrained devices. Through the exploration of network parameters such as depth, width, and resolution, it is possible to reduce the number of parameters and floating point operations (FLOPs) while preserving accuracy, as demonstrated in the works of SqueezeNet \cite{iandola2016squeezenet}, GoogleLeNet \cite{szegedy2015going}, and EfficientNet \cite{tan2019efficientnet}. Additionally, MobileNets \cite{howard2017mobilenets} pioneers the use of depth-wise separable convolutions to reduce the number of convolution operations, which has since become a fundamental component in subsequent studies.


\textbf{Adaptive Sampling}. Another effective approach is to conserve power through sparse yet strategic samplings. The approach known as adaptive sampling, which involves guiding the sampling process in a sequential manner based on information from previous observations \cite{castro2006compressed}. This allows for more efficient data collection by focusing resources on areas of interest or significance, while reducing the sampling rate in less informative regions. Superior performance compared to conventional (uniform) sampling schemes has been demonstrated using either filter-based approaches \cite{bishop2001adaptive}\cite{jain2004adaptive} or recent learning-based approaches \cite{lookman2019active}\cite{huo2020adaptive}. Adaptiveness in data sampling can be determined by considering criteria such as information gain \cite{lookman2019active} event detection \cite{qi2013adasense}, and uncertainties reduction \cite{lermusiaux2016science}. Signal reconstruction from adaptive sampling has also been investigated \cite{alippi2007adaptive}\cite{gadde2014active}. 

However, our laboratory experiments and the findings presented in \cite{kang2021benchmarking}, utilizing the Google Coral Dev Board, demonstrate that no existing lightweight object detection neural networks can sustain continuous video processing at a frequency of 30 Hz for more than 48 hours on a 10,000 mAh Li-ion battery, when deployed at a plaza with moderate pedestrian density. It is worth noting that while experiments in \cite{casares2011adaptive} demonstrate the potential for energy savings through strategic sensor activation, there is currently \textit{no existing research specifically addressing data collection using adaptive sampling in the transportation domain}.
 
Acknowledging the limitations and gaps in the existing literature, we introduce a modularized framework that is independent of specific video processing methods and offers flexibility to incorporate various adaptive sampling algorithms. Furthermore, the accuracy of data collection is enhanced through post-estimation techniques. This framework is suitable for various types of traffic data and is designed to operate on low-cost lightweight devices with limited battery power resources, like Google Coral Dev Board and Nvidia Jetson Nano.

\section{Methodology}
\label{section: III}
In this section, we delve into the proposed solution, including problem definition, framework architecture, and the design of the framework's functionality. Throughout the entire paper, we will treat sampling as making video-based observations from deep neural networks (e.g., pedestrian detection and counting) for the sake of simplicity. 

\begin{figure}[ht]
    \centering

    \includegraphics[width=1\columnwidth]{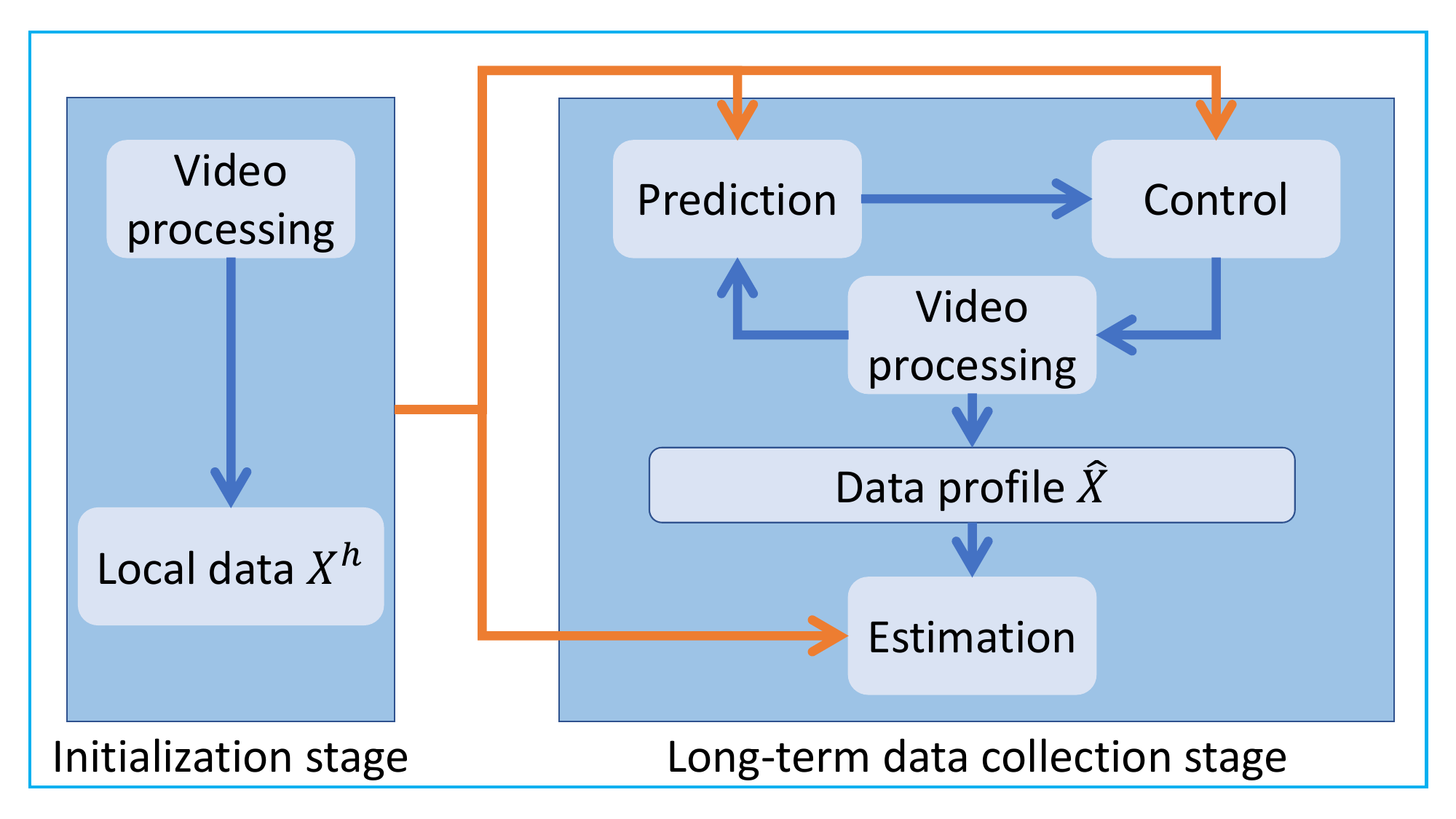}
    \caption{Workflow of the proposed two-stage framework.}
    \label{fig: workflow}
    \vspace{-4mm}
\end{figure}

\subsection{Problem statement}
 This paper addresses the challenge of optimizing the timing of observations for a lightweight device with limited battery capacity. We assume that the number of available observations is proportional to the remaining battery power. The goal is to allocate observation opportunities in a manner that maximizes the capture of information and enables the reconstruction of periods where no observations are made.

We design a two-stage framework. The first stage, referred to as the "initialization" phase, involves operating the device to gather continuous observations within a limited time period solely for the recording purpose, without any background algorithm running. The length of this stage is exclusively determined by the battery capacity and remains uninfluenced by the particular scenes encountered. In the second stage, depending on the remaining battery capacity, the device can either be equipped with a new battery or continue using the existing one. During this stage, the device transitions into "power-saving" mode, strategically making observations using the previously collected data to extend its lifespan. The workflow of the proposed framework in this work can be summarized as depicted in Fig. \ref{fig: workflow}.



\subsubsection{Prediction}
The objective of the prediction module is to forecast the future based on past observations. Let $\mathcal{P}$ denotes a predictor, $X^t$ represents the data observed at time $t$, and $\bar{X}^{t+k}$ denotes the prediction at time $t+k$. A prediction task maps $K'$ historical data to $K$ future data is given by Eq. (\ref{predictor}). 
\begin{equation}
\label{predictor}
    \begin{aligned}
       [X^{t-K'+1}, \cdots, X^t] \xrightarrow{\mathcal{P}} [\bar{X}^{t+1}, \cdots, \bar{X}^{t+K}]
    \end{aligned}
\end{equation}

In the context of time progression, accurate prediction is essential as the available information can become outdated and biased. This presents challenges for the controller in making informed decisions. To address this, we utilize an RNN predictor that is capable of capturing location-specific patterns, enhancing its sensitivity to the unique characteristics of each location. By incorporating this predictor into our framework, we aim to improve the accuracy and reliability of decision-making processes. Let $D$ denote the existing database generated from infrastructures, and $\mathbf{X^h}$ represent the data collected by the device on-site in the first stage. We apply conditional training and inference, incorporating $\mathbf{X^h}$ as part of the input, as illustrated in Eq.
(\ref{predictor_conditioned}). 
\begin{equation}
\label{predictor_conditioned}
    \begin{aligned}
        [X^t, \mathbf{X^h}] \xrightarrow{\mathcal{P}^D } [\bar{X}^{t+1}, \cdots, \bar{X}^{t+K}]
    \end{aligned}
\end{equation}

In this manner, the predictor can utilize both the existing database $D$ and the local historical data $\mathbf{X^h}$, allowing it to effectively handle limited on-site data while considering similarities across locations. We train the RNN predictor $f_{\boldsymbol{\theta^{pred}}}$, parameterized by $\boldsymbol{\theta^{pred}}$, to predict the entire future data profile using $\mathbf{X^h}$ as input. The prediction model is trained using the loss function shown in Eq. \ref{predictor_loss}, where $Y$ represents the ground truth data to be collected.
\begin{equation}
\label{predictor_loss}
    \begin{aligned}
        \mathcal{L}(\boldsymbol{\theta^{pred}}) = {\big|\big|(Y - f_{\boldsymbol{\theta^{pred}}}(\mathbf{X^h} )\big|\big|}^2_2
    \end{aligned}
\end{equation}

\subsubsection{Control}

 In this study, the selection of observation times is treated as a sequential decision-making problem, as future state-action pairs are influenced by past decisions. To incorporate temporal dependencies and find an observation policy, we employ a Deep Recurrent Q-Network (DRQN), which combines a Long Short-Term Memory (LSTM) \cite{hochreiter1997long} and a Deep Q-Network, to learn the Q-value function by integrating information from sequential inputs over time. Reinforcement Learning \cite{sutton1998introduction} is generally defined as a Markov Decision Process (MDP) problem, denoted by a tuple $\{\mathcal{S}, \mathcal{A}, \mathcal{P}, \mathcal{R}\}$. At each step, the agent observes a state $s \in \mathcal{S}$, selects an action $a \in \mathcal{A}$, and observes the next state $s' \in \mathcal{S}$. The reward is determined by the function $\mathcal{R}(s,a,s')$, which assigns a reward value to the state-action transition. In the DRQN approach, the Q-function $q_{\pi}(s,a) \approx q_{\pi}(s,a;\boldsymbol{\theta})$ is approximated using neural networks with parameters $\boldsymbol{\theta}$. The network parameters are optimized by minimizing the loss function defined in Eq. \ref{DRQN_loss}.
\begin{equation}
\label{DRQN_loss}
    \begin{aligned}
        \mathcal{L(\boldsymbol{\theta})} = \mathbb{E}_{(s,a,r,s')\sim \mathcal{B}} \Big[\big( y- Q(s,a;\boldsymbol{\theta}) \big) ^2  \Big]
    \end{aligned}
\end{equation}

Here $\mathcal{B}$ denotes the replay memory buffer containing experience $(s,a,r,s')$, and $y = r + \gamma \max_{a'}\hat{Q}(s',a';\boldsymbol{\theta^{-}})$, where $\hat{Q}(s',a';\boldsymbol{\theta^{-}})$ is the output of the target network and $Q(s,a;\boldsymbol{\theta})$ is the output of the evaluation network.

Our design incorporates the principles of Model Predictive Control (MPC) \cite{garcia1989model}, although we use a neural network for prediction instead of a dynamics model. The concept of conditional training, which is employed in the prediction module, is also applied here. The detailed design is outlined as follows:

\begin{itemize}
    \item \textit{\textbf{State}}: $s = [X^t, \bar{X}^{(t+1):(t+K)}, t_{local}, t_{global}, O_{ava}, \mathbf{X^h}]$, represents the data collected at time $t$ through observations, $\bar{X}^{(t+1):(t+K)}$ denotes the prediction from time $t+1$ to $t+K$, $t_{local}$ indicates the index of the current time $t$ in each run, $t_{global}$ corresponds to the 24-hour time in the real world, $O_{ava}$ denotes the remaining observations, and $\mathbf{X^h}$ represents the local historical data collected in the first stage. 
    \item \textit{\textbf{Action}}: $a \in \{1,2,\cdots, K\}$,
    represents the number of time steps from the current time $t$ to the next observation. This choice of action results in a finite action space, while still providing the flexibility to handle long time horizons.
    \item \textit{\textbf{Reward}}: $r(s,a,s') =  -r_{accuracy} -w_1 \cdot r_{similarity} - w_2 \cdot r_{waste}$, where $r_{accuracy} $ quantifies the prediction accuracy from time $t$ to the next observation at time $t+a$, $r_{similarity}$ measures the similarity between the prediction and the ground truth, $r_{waste}$ penalizes the agent for leaving observations unused, and weights $w_1$ and $w_2$ are weights that determine the importance of each component of the reward.
\end{itemize}

By employing DRQN, the controller evaluates a sequence of states leading up to time $t$, enabling it to ascertain the best timing for the subsequent observation.

\subsubsection{Estimation}
When the device terminates due to running out of available observations, the collected data profile consists of a combination of observed values and predictions between two consecutive observations, denoted as $\mathbf{\hat{X}} = [\cdots, X^{a_i}, \bar{X}^{(a_i+1):(a_i + a_{i+1}-1)}, X^{a_i+a_{i+1}},\cdots]$, where $a_i$ represents the action taken at each decision iteration until using up all $\mathcal{O}$ observation opportunities, with $i = 1, 2, \cdots, \mathcal{O}$. Given the ground truth of the data to be collected, the estimation becomes a supervised learning problem aimed at calibrating the prediction using the actual observed data. In this problem, the input to the estimation model is the data collected by the device along with the predicted values between the observed data points, and the label is the corresponding ground truth. We train another RNN estimator $f_{\boldsymbol{\theta ^{est}}}$, parameterized by $\boldsymbol{\theta^{est}}$, to perform post estimation with the collected data. The network is trained using the loss function shown in Eq. \ref{estimator_loss}, where $Y$ represents the ground truth of the data to be collected.
\begin{equation}
\label{estimator_loss}
    \begin{aligned}
        \mathcal{L}(\boldsymbol{\theta^{est}}) = \big|\big|Y - \mathbf{\hat{X}} \big|\big|_2^2
    \end{aligned}
\end{equation}

\subsection{Summary}

In this framework, each module plays a distinct role. The prediction module generates predictions that are incorporated as part of the state input for the control module. The control module utilizes this information to make decisions on observation times. Finally, the estimation module calibrates the data profile obtained from the control and prediction modules. Specifically, we employ a predictor based on the RNN architecture, utilizing fully connected LSTM hidden units. The design of the predictor follows the Encoder-decoder framework introduced in \cite{sutskever2014sequence}. Our DRQN architecture is based on the design presented in \cite{hausknecht2015deep}. The estimator shares the same overall architecture as the predictor. 

The energy consumption related to neural network inference depends on how often they are invoked. In this framework, each module is limited to being called no more than once per time step, which leads to a minimal energy impact in comparison to the continuous energy drain caused by hours of camera recording.


\section{Experiment}
\label{section: IV}

While the primary focus of the proposed framework is on processing videos obtained from surveillance cameras, it is also capable of handling low-dimensional time series data processed from these videos. The framework can effectively work with various types of time series data, including occupancy, speed, flow, and more. Additionally, it is applicable to different subjects such as vehicles, pedestrians, and cyclists, making it versatile for diverse scenarios and domains.  

\subsection{Dataset}

Due to the lack of video data collected from power-constrained devices, without losing generality, we adopt a broader perspective by employing time series data obtained from the current infrastructure. We treat time series data as the processed outcomes of raw videos using computer vision algorithms, e.g., object detection. Traffic\footnote{\url{https://github.com/thuml/Autoformer}} dataset \cite{wu2021autoformer} is used for experiments, which includes hourly collected of highway occupancy rates at 861 locations from July 1, 2016, to July 2, 2018, on San Francisco Bay area freeways from PeMS. This study does not exclusively focus on or confine itself to highway occupancy data. Instead, it showcases the viability of the proposed approach for a range of traffic data collection endeavors, such as collecting pedestrian/cyclist data in urban neighborhoods. Furthermore, we intentionally retain outliers in the data to demonstrate the robustness of our models. The total 861 locations in the data source are randomly divided into three sets for training, validation, and testing using a ratio of 0.7:0.2:0.1. The 17,544 data points from each location are subsequently divided into non-overlapping sub time series, with each sub-series consisting of 216 data points (equivalent to 9-day data). We take the first 48 data points (equivalent to 2-day data) as historical data collected in the first stage and aim to estimate the occupancy rates for the subsequent 168 data points (equivalent to 7-day coverage). 

\begin{table*}[ht]
\caption{Performance comparison for traffic occupancy rate data collection. Our proposed learning-based configuration achieves the best performance in all metrics and much better generalization compatibility. ↑: the higher the better,
↓: the lower the better.}
\centering
\NiceMatrixOptions{notes/para,notes/enumitem-keys-para={itemjoin = ;\;}}
\begin{NiceTabular}{c c  c c c}
    \toprule
    \RowStyle{\bfseries}
    Metric & \multicolumn{1}{c} {Uniform+GPR} &  \multicolumn{1}{c} {Uniform+LSTM$\textbf{}_{est}$} & \multicolumn{1}{c} {AR(4)$_{kal}$+Uniform +LSTM$\textbf{}_{est}$} & \multicolumn{1}{c} {LSTM$\textbf{}_{pred}$+DRQN+LSTM$\textbf{}_{est}$ }
    \\\midrule
    RMSE (↓) \tabularnote{evaluated including missing data} &  0.0318 &  0.0223 & 0.0223 & \textbf{0.0212}\\
    MAE (↓) \tabularnote{evaluated including missing data} &  0.0176 & 0.0119 & 0.121 &  \textbf{0.0115} \\ 
    MAPE (↓) \tabularnote{excluded zero values} &  74.0\% & 60.7\% & 63.0\% &  \textbf{53.3\% (-12.20\%)} \\ 
    Coverage (↑) \tabularnote{evaluated including missing data} &  1.512 & 1.512 & 1.512 & \textbf{1.721 (+13.82\%)} \\ 
    \bottomrule
\end{NiceTabular}
\vspace{-6mm}
\label{table:metric}
\end{table*}

\subsection{Settings}

We create a demanding scenario where the device has a limited opportunity to make hourly observations, allowing it to spend at most $\frac{1}{6}$ of the total time. This corresponds to $\mathcal{O} = 168 \times \frac{1}{6} = 28$ observation opportunities over the next 7 days. This constraint enables the device to extend its lifespan by a factor of 6. Following each hourly observation captured by the onboard camera, the device performs real-time processing of the raw video data using state-of-the-art artificial intelligence techniques on-board, such as object detection. Our objective is to generate accurate descriptions of the 7-day data based on the limited observations, ensuring that the final outputs capture the underlying patterns as effectively as possible. 

\subsubsection{Predictor}
As a baseline for the prediction module, we consider a classical dynamics-based method called the Autoregressive (AR) model combined with the Kalman filter. The order selection for the AR model is determined as 4 based on autocorrelation tests. This baseline, denoted as AR(4)$_{kal}$, is widely recognized for its effectiveness in time series forecasting. Our proposed RNN predictor, denoted as LSTM$_{pred}$, comprises two recurrent layers with 128 LSTM units in both the encoder and decoder. LSTM$_{pred}$ is trained by providing the input data of the first 2 days and predicting the subsequent 7 days. We apply teacher-forcing tricks \cite{williams1989learning} during training, with an initial learning rate of $1e^{-4}$. 
\subsubsection{Controller}
We pick the uniform observation policy, where observations are taken at at equal intervals across the entire time horizon, as the baseline control policy against the proposed DRQN. In this study, we construct a DRQN network consisting of three fully connected layers, followed by a recurrent layer, and another fully connected layer that maps the hidden states to the action space. We set the prediction window $K = 12$, length of history state-action pair stored in the memory replay buffer $L_{m} = 12$, and the size of replay buffer $\mathcal{B}$ is 5,000. The exploration rate $\epsilon$ decays from 1 to 0.1. The reward function calculates $r_{accuracy}= \frac{1}{a} \sum^a_{i=1}\|\bar{X}^{t+i}_{pred} - X^{t+i}_{gt}\|_2$, and $r_{similarity} = DTW(\bar{X}^{t+1:t+a}, X^{t+1:t+a})_{gt}$ is calculated using Dynamic Time Warping \cite{muller2007dynamic} implemented by \textit{tslearn} package. The last term for unused observations is calculated as $r_{unused} = \mathcal{O} - O^{T}_{ava}$, where $\mathcal{O}$ is the total available number of observation hours at the beginning and $O^{T}_{ava}$ is the number observation hours available at the end of the desired lifespan $T$. Additionally, we set $w_1 = 1$ and $w_2 = 10$.

\subsubsection{Estimator}
Two estimation methods are considered in the experiments: (1) Gaussian process regression (GPR) implemented using \textit{SciPy} package; (2) Recurrent Neural Network, LSTM$_{est}$, which shares the same architecture as LSTM$_{pred}$ but with an input layer dimension of 216 (48+168) and hidden layers of 256.

\subsubsection{Baselines}
To evaluate the effect of three modules, we put forward three baseline configurations, including (1) Uniform + GPR: Uniform observation policy and GPR estimator; (2) Uniform + LSTM$_{est}$: Uniform observation policy and the LSTM-based estimator; (3) AR(4)$_{kal}$ + Uniform + LSTM$_{est}$: the predictor is an Auto-Regressive model with a Kalman filter, combined with the Uniform observation policy and the LSTM-based estimator. We will compare the baseline configurations with our proposed configuration of LSTM$_{pred}$ + DRQN + LSTM$_{est}$. Input data for all modules in all experiments are normalized in the same magnitude.

\subsection{Long-term data collection performance comparison}

Table \ref{table:metric} shows the comparison of different configurations. They are evaluated based on three commonly used metrics for accuracy metrics, as well as a novel metric that is specifically relevant to the transportation community, including: (1) Mean Absolute Error (MAE), (2) Mean Absolute Percentage Error (MAPE), and (3) Root Mean Squared Error (RMSE), and the last metric (4) Coverage: this metric represents the cumulative sum of all observed values, reflecting the intuition that a higher quantity of content in observations can potentially yield more valuable information. All data, including missing (zeros) values, are used in calculating these metrics, with the exception of MAPE for numerical reasons. 

\begin{figure}[htbp]
    \centering
    \vspace{-4mm}
    \includegraphics[width=\columnwidth]{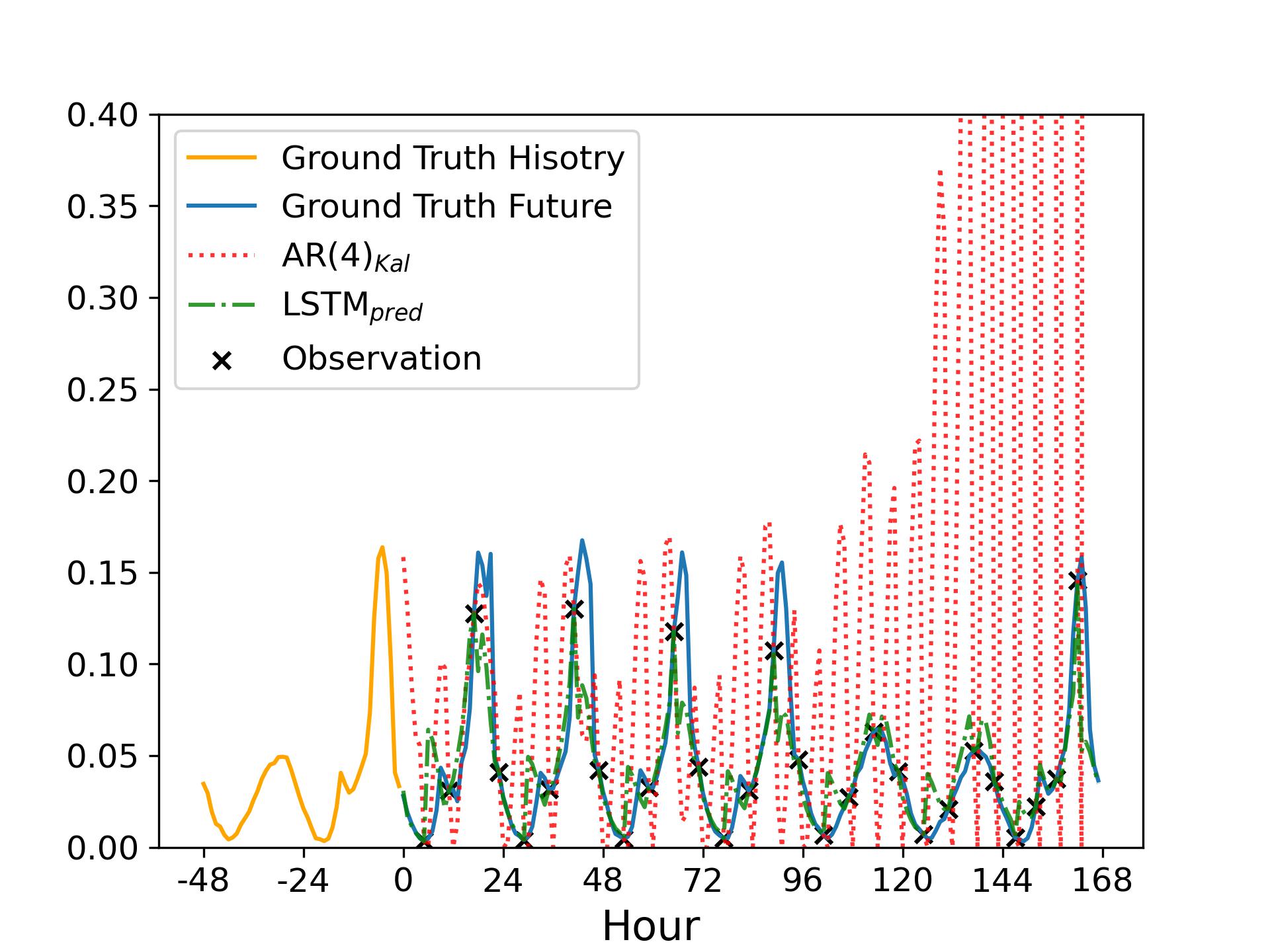}
    \caption{Prediction of AR(4)$_{kal}$ and LSTM$_{pred}$ following Uniform observation policy. LSTM$_{pred}$ can work with historical data that will fail AR(4)$_{kal}$.}
    \label{fig:AR_LSTM}
    \vspace{-4mm}
\end{figure}

\begin{figure*}[ht]
    \centering
    \includegraphics[width=1\textwidth]{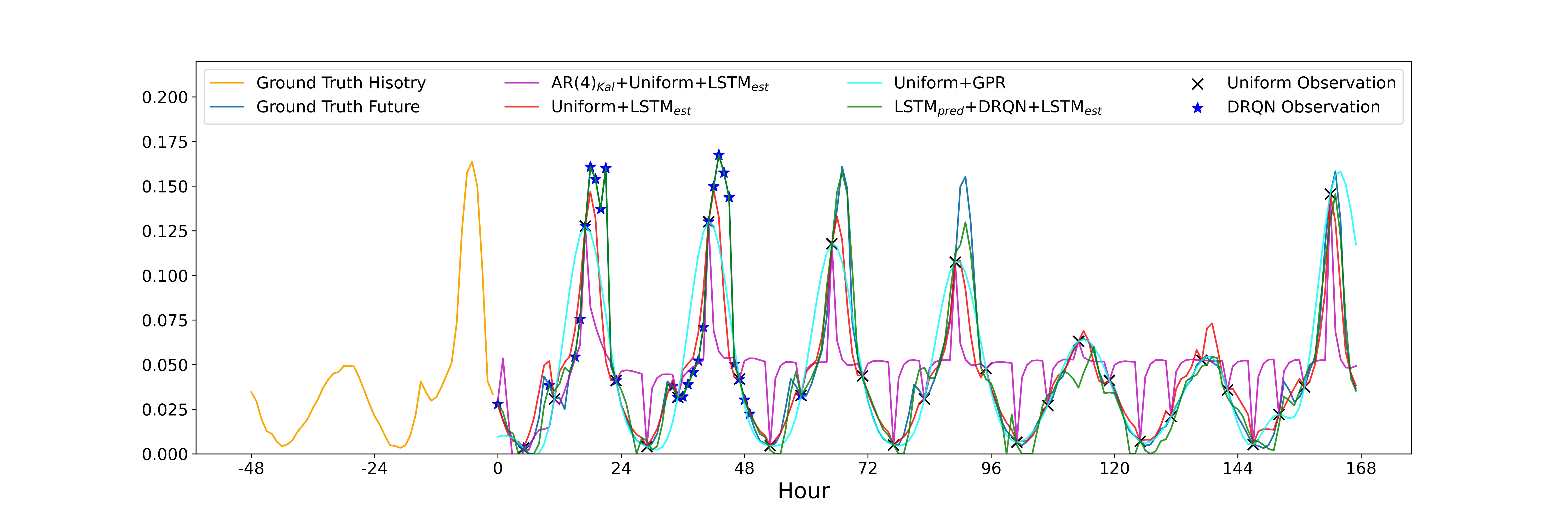}
    \caption{The proposed {\color{ForestGreen}LSTM$\textbf{}_{pred}$+DRQN+LSTM$\textbf{}_{est}$} configuration predicts the start and end of the peak hours. It can generate smooth predictions even without new observations.}
    \vspace{-8mm}
    \label{fig:metrics}
\end{figure*}

Our proposed configuration consistently outperforms the other configurations across all metrics, indicating the effectiveness of our framework and the chosen configuration (Fig. \ref{fig:metrics}). Throughout the experiments, we observed that although AR(4)$_{kal}$ generally performs satisfactorily, it can encounter significant errors or even fail in certain instances where the historical data used to fit the AR model deviates significantly from the actual data. This discrepancy can lead to exaggerated errors and ultimately cause the DRQN controller to fail due to numerical issues in loss backpropagation. Fig. \ref{fig:AR_LSTM} demonstrates a specific example where the combination of AR(4)$_{kal}$ and the Uniform policy fails to produce meaningful results. This type of failure is not acceptable in the context of long-term data collection, regardless of its low probability, as real-world data cannot always be assumed to be stationary. In contrast, our learning-based predictor, LSTM${_{pred}}$, shows robustness in scenarios where AR(4)$_{kal}$ fails. The failure cases encountered in the AR(4)$_{kal}$ + Uniform + LSTM$\textbf{}_{est}$ configuration can provide insights into why it performs even worse than its variant without the AR(4)$_{kal}$ predictor, as these failure cases may have contaminated the estimator's parameters with extremely high loss values. This highlights the significance of the prediction module in determining the overall performance evaluation.

\subsection{Effect of control policy and estimation}

In order to better understand the impact of the control policy and the estimator, we conducted two sets of ablation studies, and the experiment settings are as follows:

\begin{itemize}
    \item LSTM$\textbf{}_{pred}$ + DRQN v.s. LSTM$\textbf{}_{pred}$ + Uniform.
    \item LSTM$\textbf{}_{pred}$ + DRQN +  LSTM$\textbf{}_{est}$ v.s. LSTM$\textbf{}_{pred}$ + Uniform + LSTM$\textbf{}_{est}$.
\end{itemize}

\begin{figure}[htp]
    \centering
    \includegraphics[width=1\columnwidth]{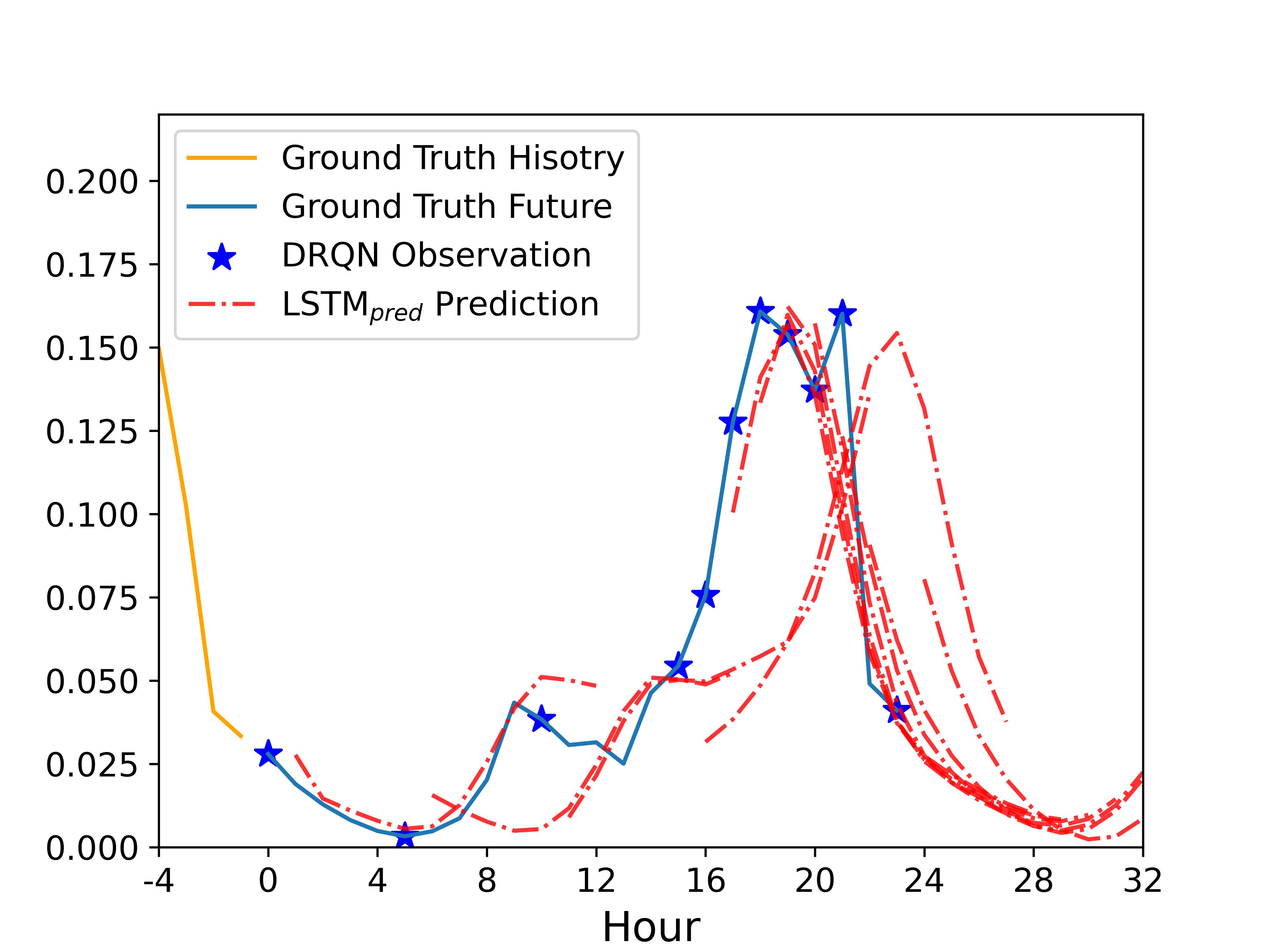}
    \caption{Observations are strategically taken either when there is a significant prediction error or at representative points, such as peaks.\\}
    \vspace{-6mm}
    \label{fig:prediction error}
    \vspace{-4mm}
\end{figure}

Table \ref{table:predictor_controller_comparision} shows the comparison between the configurations of LSTM$_{pred}$ + DRQN and LSTM$_{pred}$ + Uniform. It can be observed that DRQN results in better performance in all metrics due to its ability to actively search for the optimal observation time that minimizes the prediction error. An illustrative example depicted in Fig. \ref{fig:prediction error} shows that DRQN tends to make observations at points where the prediction errors are either significant at the current step or total deviation in the future, such as peaks or bottoms. This efficient utilization of observations helps minimize errors stemming from predictions and consequently enhances overall accuracy performance.
\vspace{-2mm}
\begin{table}[htbp]
\caption{Performance comparison for using predictor and controller only.}
\centering
\NiceMatrixOptions{notes/para,notes/enumitem-keys-para={itemjoin = ;\;}}
\begin{NiceTabular}{c c c c c}
    \toprule
    \RowStyle{\bfseries}
      &  RMSE \tabularnote{evaluated including missing data} & MAE \tabularnote{evaluated including missing data} & MAPE \tabularnote{excluded zero values} &  Coverage
    \\\midrule    
    LSTM$\textbf{}_{pred}$+DRQN&  \textbf{0.0282} & \textbf{0.0156} & \textbf{120.4\%} & \textbf{1.721} \\ 
    LSTM$\textbf{}_{pred}$+Uniform & 0.0320 & 0.0182 & 149.0\%&  1.512\\
    \bottomrule
\end{NiceTabular}
\label{table:predictor_controller_comparision}
\vspace{-4mm}
\end{table}
\vspace{-2mm}
\begin{table}[ht]
\caption{Performance comparison of DRQN and Uniform control policy given the predictor and estimator using LSTM$\textbf{}_{pred}$ and LSTM$\textbf{}_{est}$.}
\centering
\NiceMatrixOptions{notes/para,notes/enumitem-keys-para={itemjoin = ;\;}}
\begin{NiceTabular}{c c c c c}
    \toprule
    \RowStyle{\bfseries}
      &  RMSE \tabularnote{evaluated including missing data} & MAE \tabularnote{evaluated including missing data} & MAPE \tabularnote{excluded zero values} &  Coverage
    \\\midrule    
    DRQN&  \textbf{0.0212} & \textbf{0.0115} & \textbf{53.3\%} & \textbf{1.721} \\ 
    Uniform & 0.0216 & 0.0122 & 56.1\% &  1.512\\
    \bottomrule
\end{NiceTabular}
\label{table:additive_estimator_comparision}
\vspace{-1mm}
\end{table}


\begin{figure}[ht]
    \centering
    \includegraphics[width=\columnwidth]{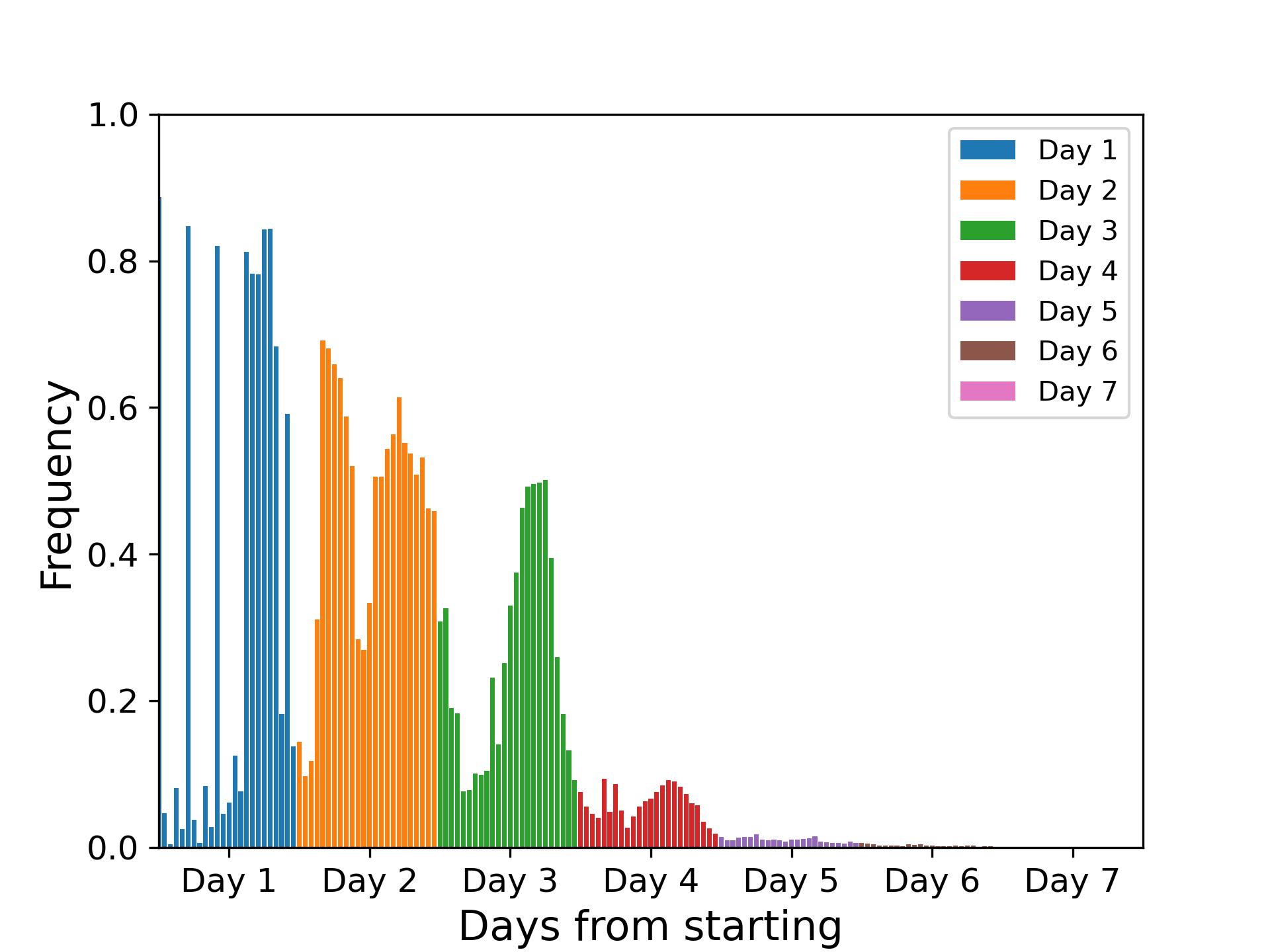}
    \caption{The distribution of observations allocated in each day by the DRQN controller. The y-axis represents the ratio of the number of observations assigned at a specific time to the total number of test instances. Each day starts at 0:00 and ends at 23:00.}
    \vspace{-6mm}
    \label{fig:action distribution}
\end{figure}

The impact of incorporating the estimation module is shown in Table \ref{table:additive_estimator_comparision}. It is observed that DRQN continues to outperform Uniform in the majority of metrics, although the performance gap narrows down, which means LSTM$\textbf{}_{pred}$ + Uniform + LSTM$\textbf{}_{est}$ gains more improvement by adding the estimator. To investigate why the addition of an estimator has a milder impact on the performance of DRQN compared to the Uniform policy, we visualize the action distribution obtained using DRQN, as depicted in Fig. \ref{fig:action distribution}. It reveals that, firstly, the allocation of observations per day decreases progressively over time, indicating that the DRQN controller assigns greater importance to the initial days compared to the later ones. This can be attributed to the fact that a general LSTM predictor has the ability to quickly learn location-specific patterns. Therefore, having more observations during the initial days facilitates faster adaptation to the local data patterns and improves the prediction accuracy for the far future. One consequence is that as the predictor becomes increasingly accurate and calibrated with the observations from the initial days, it requires fewer observations for reliable predictions. This could be one of the reasons why Uniform observations show greater improvements compared to DRQN after adding the estimator: Uniformly distributed observations provide guidance throughout the entire time horizon, while DRQN observations are concentrated in the initial days and may not contribute significantly to the estimation towards the end of the time horizon.

Secondly, the distribution of observations within each day exhibits a two-peak pattern, suggesting a higher likelihood of observations being allocated to morning and afternoon peak hours. Remarkably, the DRQN controller learns this behavior without any explicit guidance from the reward function. This demonstrates the ability of DRQN to identify crucial time intervals for maximizing performance metrics such as accuracy and information coverage, resulting in improved outcomes.  Another notable finding is that while the estimator module appears to have the most significant impact on improving accuracy performance compared to the other two modules, we believe that these three modules should function together as a cohesive system, and controllers like DRQN can bring benefits not only to accuracy but also to implicit utility in domain-specific metrics, such as information coverage in transportation.

\section{Conclusion}
\label{section: V}

This paper presents a modularized framework designed to facilitate long-term data collection on power-constrained devices. By integrating prediction, control, and estimation modules, the framework effectively extends the device's lifespan while maintaining reasonable performance. Real-world data experiments indicate the effectiveness of the proposed framework and its configuration. The effect of each module is thoroughly examined and analyzed. Future work involves testing the framework in complex urban scenarios and conducting field experiments for real-world validation.

\section*{Acknowledgment}

This work has been supported by C2SMART, a Tier 1 University Transportation Center at New York University, and the United States National Science Foundation grant \#2238968. The views presented in this paper are those of the authors alone.

\maketitle
\thispagestyle{empty}
\pagestyle{empty}




\bibliographystyle{IEEEtran}

\bibliography{reference}

\end{document}